\documentclass[prb,twocolumn,superscriptaddress,showpacs]{revtex4}
\usepackage[dvips]{graphicx}
\usepackage{latexsym}
\usepackage{bm}
\usepackage{psfrag}
\usepackage{wasysym}
\usepackage{color}
\newcommand{\fig}[2]{\includegraphics[width=#1]{#2}}
\definecolor{Blue}{rgb}{0.00, 0.00, 1.00}
\definecolor{Red}{rgb}{1.00, 0.00, 0.00}
\begin{document}
%\draft

%\newcommand{\fig}[2]{\includegraphics[width=#1]{#2}}

%\twocolumn[\hsize\textwidth\columnwidth\hsize\csname@twocolumnfalse%
%\endcsname

\newcommand{\dprime}{{\prime\prime}}
\newcommand{\be}{\begin{equation}}
\newcommand{\den}{\overline{n}} 
\newcommand{\ee}{\end{equation}}
\newcommand{\bea}{\begin{eqnarray}} 
\newcommand{\eea}{\end{eqnarray}}
\newcommand{\nn}{\nonumber} 
\newcommand{\bk}{{\bf k}}
\newcommand{\vN}{{\bf \nabla}}
\newcommand{\vA}{{\bf A}}
\newcommand{\vE}{{\bf E}}
\newcommand{\vj}{{\bf j}}
\newcommand{\vJ}{{\bf J}}
\newcommand{\bs}{{\bf S}}
\newcommand{\vn}{{\bf v}_n}
\newcommand{\vv}{{\bf v}} 
\newcommand{\la}{\langle}
\newcommand{\ra}{\rangle} 
\newcommand{\ph}{\phi} 
\newcommand{\dg}{\dagger}
\newcommand{\br}{{\bf{r}}} 
\newcommand{\bo}{{\bf{0}}} 
\newcommand{\bR}{{\bf{R}}} 
\newcommand{\bS}{{\bf{S}}} 
\newcommand{\bq}{{\bf{q}}}
\newcommand{\bQ}{{\bf{Q}}}
\newcommand{\vQ}{{\bf{Q}}} 
\newcommand{\hj}{\hat{\alpha}}
\newcommand{\hx}{\hat{\bf x}} 
\newcommand{\hy}{\hat{\bf y}}
\newcommand{\hz}{\hat{\bf z}}
\newcommand{\vS}{{\bf S}} 
\newcommand{\cV}{{\cal U}}
\newcommand{\cD}{{\cal D}} 
\newcommand{\tnh}{{\rm tanh}}
\newcommand{\sh}{{\rm sech}} 
\newcommand{\vR}{{\bf R}}
\newcommand{\crx}{c^\dg(\vr)c(\vr+\hx)}
\newcommand{\crkubox}{c^\dg(\vr)c(\vr+\hat{x})}
\newcommand{\pll}{\parallel} 
\newcommand{\crj}{c^\dg(\vr)c(\vr+\hj)}
\newcommand{\crmj}{c^\dg(\vr)c(\vr - \hj)}
\newcommand{\sumall}{\sum_{\vr}} 
\newcommand{\sumx}{\sum_{r_1}}
\newcommand{\nabj}{\nabla_\alpha \theta(\vr)} 
\newcommand{\nabx}{\nabla_1\theta(\vr)} 
\newcommand{\sumy}{\sum_{r_2,\ldots,r_d}}
\newcommand{\krj}{K(\vr,\vr+\hj)} 
\newcommand{\sigr}{|\psi_0\rangle}
\newcommand{\sigl}{\langle\psi_0 |}
\newcommand{\sier}{|\psi_{\Phi}\rangle}
\newcommand{\siel}{\langle\psi_{\Phi}|}
\newcommand{\sumrj}{\sum_{\vr,\alpha=1\ldots d}}
\newcommand{\krw}{K(\vr,\vr+\hx)} 
\newcommand{\Dtheta}{\Delta\theta}
\newcommand{\rhonew}{\hat{\rho}(\Phi)}
\newcommand{\rhoold}{\hat{\rho_0}(\Phi)} 
\newcommand{\dt}{\delta\tau}
\newcommand{\cP}{{\cal P}} 
\newcommand{\cS}{{\cal S}}
\newcommand{\vm}{{\bf m}} 
\newcommand{\hnr}{\hat{n}({\vr})}
\newcommand{\hnm}{\hat{n}({\vm})} 
\newcommand{\del}{\hat{\delta}}
\newcommand{\upa}{\uparrow} 
\newcommand{\dna}{\downarrow}
\newcommand{\dnk}{\delta n_{\vk}}
\newcommand{\dnks}{\delta n_{\vk,\sigma}}
\newcommand{\dnkp}{\delta n_{\vk '}}

\author{D. N. Sheng}
\affiliation{Department of  Physics and Astronomy, California State
University, Northridge, CA 91330}

\author{Leon Balents}
\affiliation{Department of  Physics, University of California,
Santa Barbara, CA 93106}

\title{Numerical study of fractionalization in an Easy-axis Kagome 
antiferromagnet}

\date{\today}

\begin{abstract}
  Based on exact numerical calculations, we show that the generalized
  Kagome spin model in the easy axis limit exhibits a spin liquid,
  topologically degenerate ground state over a broad range of phase
  space.  We present an (to our knowledge the first) explicit
  calculation of the gap (and dispersion) of ``vison'' excitations,
  and exponentially decaying spin and vison 2-point correlators,
  hallmarks of deconfined, fractionalized and gapped spinons.  The
  region of the spin liquid phase includes a point at which the model
  is equivalent to a Heisenberg model with purely two-spin
  interactions.  Beyond this range, a negative ``potential'' term
  tunes a first order transition to a magnetic ordered state.  The
  nature of the phase transition is also discussed in light of the low
  energy spectrum.  These results greatly expand the results and range
  of a previous study of this model in the vicinity of an exactly
  soluble point.\cite{bfg}

\typeout{polish abstract}
\end{abstract}


\maketitle

\begin{figure}[hbt]
\centerline{\fig{7cm}{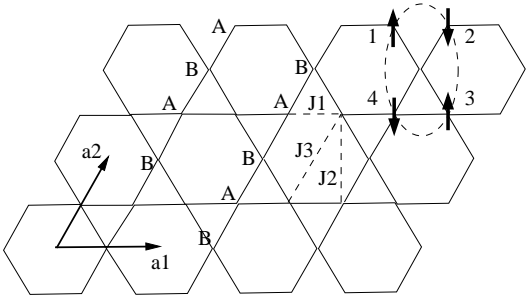}}
\caption{Kagome lattice and interactions.  Two primitive vectors
  $\vec{a}_1,\vec{a}_2$ are shown, as are the three
  2-spin couplings on sites within a Hexagon.
The ring term involves  four sites on a bow-tie, which
is generated from
2-spin virtual exchange processes. Sublattice labeling is also shown.}
\label{fig:lattice}
\end{figure}

The discovery of the fractional quantum Hall effect (FQHE) has
revealed that electronic systems can have fractionally quantized
excitations in nature.  The possibility of such fractionalization in
spin systems, without any applied magnetic field, is a subject of
great interest.  Anderson first proposed that two-dimensional (2d)
spin-1/2 antiferromagnets might condense into a featureless
``spin-liquid'' quantum ground state\cite{PWA}, with deconfined spinon
excitations carrying spin $S=1/2$\cite{RVB, RKdimer}. In the past
several years, a clear notion of fractionalized and
gapped\cite{gapless} spin-liquid states emerged.  With the absence of
spin ordering and spatial symmetry breaking, such a liquid state is
characterized by ``topological order''\cite{oldtop,newtop}, as in the
FQHE \cite{Wentop}.  Spinons are subject to long-range statistical
interactions with vortex-like excitations (denoted as visons) which
carry an Ising or $Z_2$ flux\cite{Z2, vortpair}.

%In a system with anisotropy, one can view
%such a spin-liquid state as a ``quantum disordered'' planar magnet, in
%which doubly quantized vortices -- points around which the magnetic
%order parameter winds twice -- have condensed in the ground state to
%destroy the magnetism, leaving singly quantized vortices behind as
%``visons''.  

Theoretically a few spin models have been identified as possible
candidates for realizing such a spin liquid phase\cite{bfg, kagnum,
  sondhi, Nayak, sent}.  Moessner and Sondhi\cite{sondhi} have
suggested this might occur for some antiferromagnets on the triangular
lattice, by showing that a particular so-called ``quantum dimer
model'' on the triangular lattice is in a featureless deconfined spin
liquid phase in a range of parameters.  Unfortunately, the triangular
quantum dimer model does not derive directly from a triangular lattice
spin model, so {\sl which} spin Hamiltonian might realize this state
is currently unclear.  Similar dimer models on the square lattice
display only a deconfined critical point \cite{RKdimer} separating two
confined phases.

Balents, Fisher and Girvin\cite{bfg} have recently displayed a spin
model on the Kagome lattice, closely mathematically related to the
above triangular dimer model, which demonstrably has a
fractionalized, topologically ordered ground state.  Similarly to
earlier studies on various dimer models\cite{sondhi, Nayak},
fractionalization is established through mapping the system to an
exactly soluble point, first exploited by Rokhsar and Kivelson (RK)
\cite{RKdimer} in the square lattice.  A drawback of this approach is
that the RK point requires four-spin couplings, and topological order
is argued perturbatively from the RK point.  Ref.\onlinecite{bfg}\ 
speculated, however, that the spin liquid state may persist to a
simpler limit describable by only two-spin Heisenberg interactions.
This central issue remains unsettled.

%, if fractionalization can persist
%away from the perturbation regime of the RK point.

%   At the RK point,
%the ground state can be viewed as equal weight superposition of all
%the spin configurations in the low energy singlet sector.  

In this letter, we present exact numerical diagonalization studies of
a generalized Kagome spin model\cite{bfg} in the easy axis limit (see
below) which interpolates between the RK point and a two-spin
Heisenberg form -- and beyond.  We have found that the spin liquid
phase persists over a wide range of parameters including the two-spin
limit.  This phase is characterized by the absence of spin ordering,
and by deconfined, fractionalized and gapped spinon excitations. The
spin liquid phase has four-fold topological degeneracy, a finite gap
to the vison excitations, and short range exponentially decaying spin
and vison 2-point correlators.  Still further from the RK point beyond
the 2-spin model, a first order transition to a magnetically ordered
phase occurs.  
%The detailed nature of the phase transition and the
%numerical indications for the gapless spin-liquid phase will also be
%discussed.

The model considered in most of this paper is the generalized spin-$1/2$
ring-exchange Hamiltonian:
\begin{equation}
  {\cal H} =  \sum_{\bowtie} (-J_{ring} S^+_1 S^-_2
  S^+_3 S^-_4 +  {\rm h.c.}  +  u_4 {\hat P}_{flip}), \label{eq:Sring}
\end{equation}
where the labels $1\ldots 4$ denote the four spins at the ends of each
bowtie (and others obtained by 120 degree rotations), as labeled in
Fig.~\ref{fig:lattice}.  The additional term is a projection operator,
\begin{equation}
  \label{eq:Phat}
  \hat{P}_{flip} = \left|\uparrow \downarrow \uparrow \downarrow\right\rangle
  \left\langle \uparrow \downarrow \uparrow \downarrow\right| +
  \left|\downarrow \uparrow \downarrow \uparrow\right\rangle
  \left\langle \downarrow \uparrow  \downarrow \uparrow\right|,  
\end{equation}
with spin states $S_1^z,S_2^z,S_3^z,S_4^z$ indicated sequentially in
the bras/kets.
%diagonal in $S^z_i$,
%where
%%\begin{equation}
%%  \hat{P}_{\rm flip} = \sum_{\sigma=\pm1}\prod_{j\in r =1}^4
%%  ({\frac{1}{2}} +\sigma (-1)^j S^z_j) .
%%  \end{equation}
%the operator ${\hat P}_{flip}$ is a projection operator onto the 
%flippable configurations (i.e. $\uparrow \downarrow \uparrow \downarrow$
%or $\downarrow \uparrow \downarrow \uparrow$) of the bow-tie ring.  
This Hamiltonian acts in the reduced Hilbert space with the constraint
that for each hexagon, the total $S^z$ of six spins,
$S^z_{\mbox{\small\hexagon}} =0$.  For the particular value $u_4=0$,
Eq.~(\ref{eq:Sring}) can be shown to be equivalent to the
leading-order effective Hamiltonian describing the easy-axis limit of
the Heisenberg model,
\begin{equation}
\label{eq:Heisenberg}
  {\cal H} = 2 \sum_{(ij),\mu} J^\mu_{ij} S^\mu_i S^\mu_j,
\end{equation}
where the sum is over pairs of sites $(ij)$, with non-zero
$J_{ij}^\mu=J_\mu$ for all first, second, and third nearest-neighbor
pairs $(ij)$ on the Kagome lattice (see Fig.~\ref{fig:lattice}).
Specifically, in the extreme easy-axis limit, $J_z \gg J_\perp$,
Eq.~(\ref{eq:Heisenberg}) reduces by second order degenerate
perturbation theory to Eq.~(\ref{eq:Sring}) with $J_{ring}=4
J_{\perp}/J_z^2$ and $ u_4=0$.  The energy of states with
$S^z_{\mbox{\small\hexagon}}\neq 0$ is higher by $O(J_z)$, and such
states require additional terms beyond those in Eq.~(\ref{eq:Sring}) for
their description.
                                                                                
%The operator $\hat{P}_{\rm flip}$ is a projection operator onto the two
%flippable states of the bow-tie ring $r$.  This term in the
%Hamiltonian can be combined with ${\cal H}_{ring}$ and written in the
%suggestive form:
%\begin{equation}
%  {\cal H}_{\rm ring} + {\cal H}_{\rm nf} =  \sum_{r}
%  \hat{P}_{\rm flip}(r)\lbrace -J_{\rm ring} \prod_{j=1}^4  2 S^x_j  +
%  u_4 \rbrace.
%\label{HamRK}
%\end{equation}

%We first briefly summarize the results around the exact soluble point
The exact soluble RK point corresponds to $u_4=J_{ring}$.  For
$u_4=J_{ring}-\epsilon$, with $\epsilon \ll J_{ring}$, the ground
state is a featureless spin liquid state with gaps to all excitations.
In particular, the vison gap was argued variationally to be
$O(J_{ring})$\cite{bfg}.  We have performed exact Lanczos
diagonalization in the whole range of $-2 \leq u_4 \leq 1$ (we take
$J_{ring}=1$ as the unit).  Specifically, we consider a finite size
system on the torus with length vectors $\vec{L}_1=n_1{\vec{a}_1}$ and
$\vec{L}_2=n_2{\vec{a}_2}$, which connect identical sites (periodic
boundary condition (PBC)). Here ${\vec a}_1$ and ${\vec a}_2$ are the
primitive vectors shown in Fig.~ 1, and we set lattice constant
$a_1=a_2=2$ for convenience.  Then total number of sites is
$N_s=3n_1n_2$.  In the constrained Hilbert space with
$S^z_{\mbox{\small\hexagon}} =0$, this problem is characterized by two
topological $Z_2$ ``winding numbers'', $(w_1,w_2)$
\begin{equation}
  \label{eq:winding}
  w_a =  \prod_{k}\!\!\!\!\!\!\!\!\!\!\longrightarrow  2S_k^z = \pm 1,
%\longrightarrow
\end{equation}
where, for concreteness, the product is taken on one arbitrarily
chosen straight line of sites along the $\vec{a}_1/\vec{a}_2$ axis for
$a=1,2$ encircling the torus.  These two winding numbers are complete
in that other winding numbers defined on other non-trivial loops are
not independent of these.  Using translational invariance, the Hilbert
space for $N_s=60$ can be reduced to a dimension
(total number of configurations) around $N_{conf}=3618896$ ($N_{conf}$
varies amongst different topological sectors).

\begin{figure}
\begin{center}
\vskip-2.8cm
\hspace*{-0.8cm}
\fig{2.2in}{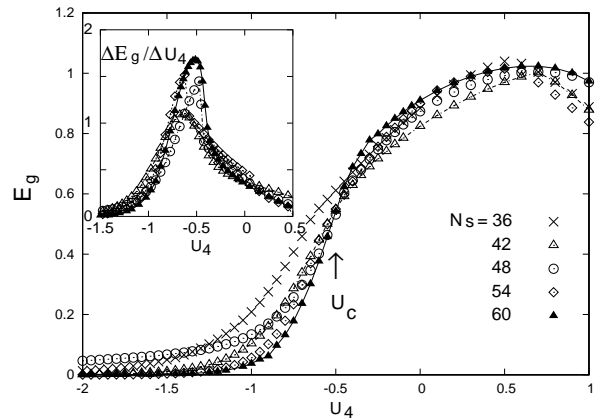}
\vskip-2mm
\caption{ 
  The spectral gap $E_{g}=E(2)-E(1)$ in the winding sector $(1,1)$ as
  a function of $u_4$ for sizes $N_s=36, 42, 48, 54$ and $60$.  In the
  inset, $\Delta E_g/\Delta u_4$ vs. $u_4$ for $N_s=42-60$ (same
  symbols as in the main plot).  }
\label{fig:fig2}
\vskip-8mm
\end{center}
\end{figure}

Shown in Fig. 2 is the excitation energy gap between the lowest two
states $E_{g}=E(2)-E(1)$ in the topological sector with
winding numbers $(1,1)$.
As $u_4$ moves away from $1$, the energy gap remains
at the order of $1$ and only becomes much smaller on the negative
$u_4$ side.  All curves actually cross over each other around
$U_c=-0.5$, and $E_g$ overall decreases with the increasing of $N_s$
with a trend of going to zero in the regime $u_4 < U_c=-0.5$.  To
examine the finite-size effect of transition, we show the slope of the
$E_g$ curve vs. $u_4$ in the inset of Fig. 2 for large sizes
$N_s=42-60$.  Indeed around $U_c=-0.5$, a strong peak showed up in
$\Delta E_g/\Delta u_4$, with its height increasing with $N_s$, which
is suggestive towards a quantum phase transition at $U_c$ to a
different phase with vanishing $E_g$.

At the RK point, a finite gap $E_g$ is expected, related to the
excitation energy of vison excitations \cite{bfg}, and as just
discussed, this gap persists for $U_c<u_4 \leq U_{RK}=1$.  Visons are
characterized by the $Z_2$ flux $\Phi_\pi=\pm 1$, which is defined on
each triangular plaquette of the Kagome lattice, as\cite{bfg}
\begin{equation}
  \Phi_\triangle = \prod_{\bowtie\in \triangle} \prod_{i\in\bowtie}2 S_i^x,
\end{equation}
where the first product is over the three bowties centered on the
sites of the triangle, and the second is over the sites of each
bowtie.  It is simple to show that a ground state $|0\rangle$ of
${\cal H}$ can always be found which is expressed as a superposition
of ${S_i^z}$ eigenstates with {\sl positive real} coefficients.  This
implies that $\langle 0| \Phi_\triangle|0\rangle >0$; hence, there are
no visons in the ground state.  One can readily show that with PBCs
the product of $\Phi_\triangle$ over all triangular plaquettes (even
over just all say up-pointing triangles) is $+1$, hence visons
($\Phi_\triangle=-1$) can only appear in pairs with PBCs for ${\cal
  H}$.  An appropriate definition of a single vison state is made as
follows.  We imagine a large open system, and perform the canonical
transformation
\begin{eqnarray}
\label{eq:hpdef}
  {\cal H}'  & = & \hat{v}_{i0} {\cal H} \hat{v}_{i0}, 
\end{eqnarray}
where $\hat{v}_{i0}$ denotes a single vison creation operator
identified in Ref.~\onlinecite{bfg}, which is a ``string'' operator
made of the product of spins $2S^z_i$ along some path on the kagome
lattice starting at site $i_0$ and ending at the boundary.
Explicitly, ${\cal H'}$ is the same as the ${\cal H}$ except that the
ring exchange terms of the three bow-ties centered on the triangle
containing $i_0$ from which the path exits are changed in sign.  This canonical
transformation {\sl redefines} the $Z_2$ flux on this triangle only by
a minus sign, $\Phi_\triangle \rightarrow -\Phi_\triangle$, i.e. a
vison on this triangle corresponds now to $\Phi_\triangle=+1$.
Imposing PBCs on ${\cal H}'$ (it is no longer canonically conjugate to
${\cal H}$) then {\sl forces} the total number of visons to be odd.
Although it is not manifest, ${\cal H}'$ has a hidden ``magnetic''
translational invariance, so there is no localization of the forced
vison.  Thus the energy difference between the ground states of the
two Hamiltonians, $E_{sv}=E'(0) -E(0)$ gives the single vison energy.
We plot $E_{sv}$ vs. $u_4$ in Fig. 3a for system sizes $N_s=36-60$.

\begin{figure}
\begin{center}
\vskip-2.8cm
\hspace*{-1.8cm}
\fig{2.2in}{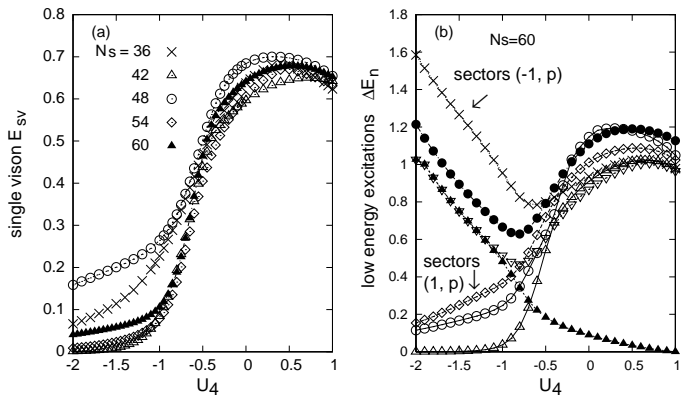}
\vskip-2mm
\caption{ 
(a) The single vison energy $E_{sv}=E'(0) -E(0)$ as a function of $u_4$
for system sizes $N_s=36, 42, 48, 54$ and $60$; (b) The low energy 
spectrum $\Delta E_n =E(n)-E_{ground}$ for $N_s=60$ 
($L_1=5a_1$ and $L_2=4a_2$), in all four sectors vs.  $u_4$.}
\label{fig:fig3a}
\vskip-8mm
\label{fig:fig4}
\end{center}
\end{figure}

$E_{sv}$ shows overall similar behavior to the spectral gap $E_g$, but
bigger than 0.5$E_{g}$ in the vison gapped regime, indicating a non-vanishing
binding energy between two visons.
%in the first excited state of the uniform ${\cal H}$, which is about $0.3$.
In the vison gapped regime, we find that the low energy manifold of
${\cal H'}$ forms an energy band of single visons with very
small dispersion (eg., for $N_s=60$ dispersion of single vison energy
is about $0.047J_{ring}$ at $u_4=0$).
%while the ground state of ${\cal H}$ is nondegenerate in each 
% topological sector.
$E_{sv}$ remains to almost constant in the regime passing $u_4=0$, and
drops significantly at $u_4<U_c=-0.5$ side \cite{note}.

To reveal the nature of the phase transition at $U_c$, we present the
low energy spectrum for $N_s=60$ in all four topological sectors by
plotting $\Delta E_n=E(n) -E_{ground}$ vs. $u_4$ in Fig. 3b (where
$E_{ground}$ refers to the lowest state energy of the system).
Clearly, in the whole range of $u_4>U_c$, the lowest energy states
from each topological sector remain quasidegenerate (filled
triangular-up, is the energy difference).
%Due to $n_1=5$ which is odd, every two topological sectors with different
%winding numbers along ${\vec L}_2$ are identical\cite).
%This  demonstrates that the present  spin state  has 
%no long-range spin ordering but  topological ordering.
At $u_4 < U_c$, the states with different winding numbers along ${\vec
  L}_1$, are separated into two groups.  All the low energy states
from the sectors $(1,\pm 1)$ go down and collapse with the ground
state. The other $(-1,\pm 1)$ states rise and become well separated
from the low energy manifold. This dependence on the topological
sector is evidence of developing spin long-range ordering.  Similar
results are obtained for all system sizes $36-66$.
% (when both $n_1$ and $n_2$ are even, the low energy states are 
% entirely in the (1,1) sector).  
The phase transition between the vison gapped state and spin
ordered phase seems likely first order as $E_g$ crosses at $U_c$ and
becomes zero at $u_4<U_c$ side. At almost the same time, the ground
state degeneracy of all sectors breaks done and long-range spin order
develops.

As discussed in Ref.~\onlinecite{Z2}, exponential decay of the
vison-vison correlation function is the hallmark of a 2d $Z_2$
fractionalized phase\cite{Z2}.  Following Ref. ~\onlinecite{bfg}, we
define the spin and the vison 2-point correlation functions:
\begin{equation}
C_{ij} = |\langle 0| S_i^z S_j^z |0\rangle|,\!\hspace{0.4in}
 V_{ij} = \left|\langle 0| 
    \prod_{k=i}^{j}\!\!\!\!\!\!\!\!\!\!\longrightarrow  2S_k^z
    |0\rangle\right|,  \label{eq:vij}
\end{equation}
where $|0\rangle$ denotes the ground state, and the product in
$V_{ij}$ is taken along some path on the Kagome lattice starting at
site $i$ and ending at site $j$, containing an even number of sites,
and making only ``$\pm 60^\circ$'' turns left or right.  Due to the
constraint $S^z_{\mbox{\small\hexagon}} =0$, the latter product is
path-independent up to an overall sign (hence the absolute value in
Eq.~\ref{eq:vij}).

To examine the longer distance behavior, we considered a strip
geometry with ${\vec L}_2= 2 {\vec a}_2$, while varying
${\vec L}_1=n_1 {\vec a}_1$.  Similar analysis of the
spectral gap $E_g$ and single vison energy $E_{sv}$ reveals that the
critical $U_c$ for strip-like samples is slightly more negative than
that for more two-dimensional clusters, around $-0.75$.

\begin{figure}
\begin{center}
\vskip-2.8cm
\hspace*{-2.8cm}
\fig{2.6in}{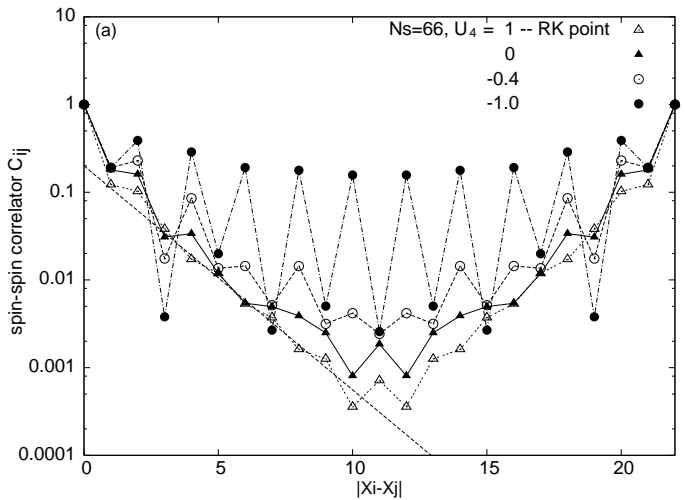}
\caption{ 
The spin correlator $C_{ij}$  vs. distance
$x_i-x_j$ at $u_4=1,0,-0.4$, and $-1$ for strip-like
system at $N_s=66$. 
}
\label{fig:fig4}
\vskip-8mm
\end{center}
\end{figure}

The numerically calculated $C_{ij}$ is shown in Fig. 4. for $N_s=66$
and different $u_4$.  At $u_4=1 $ (RK-point), $C_{ij}$ clearly shows
the exponential decay seen previously from the exact
wavefunction\cite{bfg}.  As $u_4 $ varies from 1 to 0, apart from a
small oscillation, we see essentially the same exponential behavior.
The data at $u_4=1$ and $0$, can both be well fitted by $\ln C_{ij}
\sim -|x_i-x_j|/\xi$ with apparently the same correlation length $\xi
\approx 1.7$.  For $u_4$ further decreased to $-0.4$, just before the
phase transition, much stronger fluctuations emerge between $x_i-x_j$
even or odd, but the overall decay remains exponential.  At $u_4=-1$,
$C_{ij}$ jumps by more than one order of magnitude at longer distance
$x_i-x_j=L_1/2=11$, and a longer range correlation is clearly evident.

\begin{figure}
\begin{center}
\vspace*{0.5cm}
\hspace*{-1cm}
\fig{2.in}{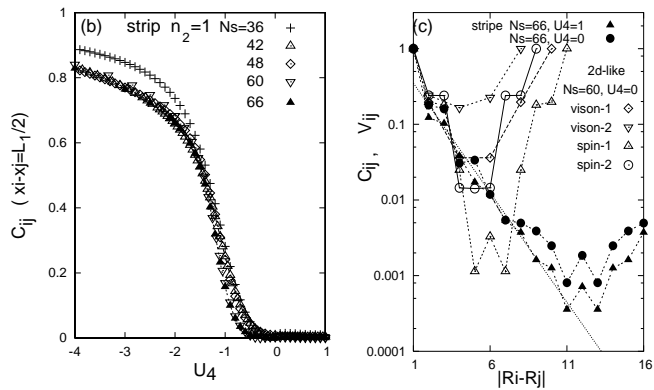}
\vskip-0.2cm
\caption{
(a) $C_{ij} $ at fixed $x_i-x_j=L_1/2$ vs. $u_4$ for strip-like
system at $N_s=36, 42, 48, 60, 66$.
(b) Vison correlator $V_{ij}$ along ${\vec a}_1$  (vison-1)
and ${\vec a}_2$ (vison-2) for 2d-like system $N_s=60$ 
are compared with $C_{ij} $  for both 2d system at $N_s=60$ and
stripe system at $ N_s=66$.  }
\label{fig:fig5}
\vskip-8mm
\end{center}
\end{figure}

We analyze the finite-size dependence of the spin spin correlation
function.  In Fig. 5a, $C_{ij}$ with $j=i+L_1/2$ (halfway across the
torus) is shown for relatively large sizes $N_s=36,42,48,54,60$ and
$66$.  For $u_4>U_c$ (around -0.75), $C_{ij}$ is vanishingly small for
all sizes.  For $u_4<U_c$, however, $C_{ij}$ is very weakly dependent
on $N_s$ (except for the smallest system with $N_s=36$), and it
robustly scales to a finite value at large $N_s$ limit.  This strongly
indicates long-range magnetic order.  One expects this region is
adiabatically connected to the $u_4 \rightarrow -\infty$ limit, for
which the ground state is fully magnetically ordered state with
$4\langle S_iS_j \rangle=\pm 1$, taking the positive or negative sign
if $i$ and $j$ belong to the same or different sublattices,
respectively (see sublattice A and B labeling in Fig. 1; spins not
belonging to these two sublattices are not ordered).
% {\bf We need to change the figure to actually show sublattices}
%Very similar results are obtained for clusters with $L_1$ and $L_2$
%close to each other (2d-like clusters).
%The nature of the magnetic ordered phase depends on  the
%dimension of the clusters.  
In the strip geometry for large negative $u_4$, one can show the
ground state manifold (with 4-fold degeneracy) is well separated from
excited states by an energy gap of $4J_{ring}/|u_4|^2$, with
additional ordering in the x-y plane (in spin space), as a result of
order-by-disorder phenomena.  Details will be reported elsewhere.

We further examine the vison 2-point correlator.
%, in both the strips and more isotropic clusters.  
In the strip case, the vison correlator
vanishes exponentially but with an extremely short correlation length
$\xi$ much smaller than the lattice constant.  For more 2d clusters,
the vison correlator has almost the same $\xi$ as the spin correlator.
We plot $C_{ij}$ and $V_{ij}$ at $u_4=0$ for cluster of $N_s=60$, with
$L_1=5a_1$ and $L_2=4a_2$, vs. $R_i-R_j$ (along both ${\vec a}_1$ or
${\vec a}_2$ as vison-1 or  vison-2) in Fig. 5b.  Despite the very short 
distance across the torus, the several points indeed follow an exponential
behavior comparable with $C_{ij}$ for the strip with $N_s=66$.
Clearly, all the correlators can be well fitted by $\exp(
-|R_i-R_j|/\xi)$ (as the dashed line in the plot) with $\xi \approx
1.7$.  

We conclude with some discussion of related quantum phase transitions.
Our numerical calculations indicate that the fractionalized
spin-liquid state, previously established for the generalized Kagome
spin model at the RK point, persists in a wide range of parameters,
covering the pure 2-spin Kagome model ($u_4=0$).  On the negative $u_4$ side,
a first order transition to the magnetic ordered phase happens at
$U_c= -0.5 J_{ring}$.  It seems
that the easy axis spin model easily develops long-range Ising
magnetic order at the confinement transition.  It is possible that
this ordering is an artifact of the easy-axis limit, which suppresses
some fluctuations, and perhaps Kagome antiferromagnets away from
easy-axis limit may develop instead a gapless spin-liquid phase, which
would be extremely interesting to explore.

We would like to acknowledge stimulating discussions with Matthew
Fisher, and both stimulation and patient explanation from Arun
Paramekanti.  This work was supported by ACS-PRF  36965-AC5 and
41752-AC10, Research Corporation Award CC5643, the NSF  grants
DMR-0307170 (DNS), DMR-9985255, the Sloan and
Packard foundation (LB).  
We  thank the  Aspen Center for Physics and
Kavli Institute for Theoretical
Physics for hospitality and support 
(through PHY99-07949 from KITP),
where part of this work was done.


\begin{thebibliography}{99}
\bibitem{gapless}  Other gapless spin-liquid states are also possible, but
less well-understood.
\bibitem{note} In the large negative $u_4$ case, $E_{sv}$ may not scale
to zero, which then corresponds to the energy cost of frustrating
the original uniform system.
\end{thebibliography}
\end{document}